\documentclass[times]{aa}
\usepackage{graphicx}
\def\folio{\ifnum\pageno=1\nopagenumbers\else\number\pageno\fi}

\def\lax    {\ifmmode{_<\atop^{\sim}}\else{${_<\atop^{\sim}}$}\fi}
\def\gax    {\ifmmode{_>\atop^{\sim}}\else{${_>\atop^{\sim}}$}\fi}
\newbox\grsign      \setbox\grsign=\hbox{$>$} 
\newdimen\grdimen   \grdimen=\ht\grsign
\newbox\simgreatbox \setbox\simgreatbox=\hbox{\raise.5ex\hbox{$>$}\llap
                        {\lower.5ex\hbox{$\sim$}}}\ht1=\grdimen\dp1=0pt
\newbox\simlessbox  \setbox\simlessbox =\hbox{\raise.5ex\hbox{$<$}\llap
                        {\lower.5ex\hbox{$\sim$}}}\ht2=\grdimen\dp2=0pt
\def\simgreat{\mathrel{\copy\simgreatbox}}
\def\simless {\mathrel{\copy\simlessbox }}
\def\pz {\phantom{$>$}}
\newbox\grsign \setbox\grsign=\hbox{$>$} \newdimen\grdimen \grdimen=\ht\grsign
\newbox\laxbox \newbox\gaxbox
\setbox\gaxbox=\hbox{\raise.5ex\hbox{$>$}\llap
     {\lower.5ex\hbox{$\sim$}}}\ht1=\grdimen\dp1=0pt
\setbox\laxbox=\hbox{\raise.5ex\hbox{$<$}\llap
     {\lower.5ex\hbox{$\sim$}}}\ht2=\grdimen\dp2=0pt
\def\gax{\mathrel{\copy\gaxbox}}
\def\lax{\mathrel{\copy\laxbox}}
\def\boxit#1    {\vbox{\hrule\hbox{\vrule\kern3pt
                  \vbox{\kern3pt#1\kern3pt}\kern3pt\vrule}\hrule}}
\def\h      {\ifmmode{^{\rm h}}\else{$^{\rm h}$}\fi}
\def\m      {\ifmmode{^{\rm m}}\else{$^{\rm m}$}\fi}
\def\s      {\ifmmode{^{\rm s}}\else{$^{\rm s}$}\fi}
\def\decas    {\ifmmode{{\rlap.}{''}}\else{${\rlap.}{''}$}\fi}
\def\mum     {\ifmmode{\mu{\rm m}}\else{$\mu{\rm m}$}\fi}
\def\s      {\ifmmode{^{\rm s}}\else{$^{\rm s}$}\fi}
\def\decdeg {\rlap . {}^\circ}     %e.g. $40\decdeg 5$ for 40.5 degrees
\def\deg      {\ifmmode{^{\circ}}\else{$^{\circ}$}\fi}
\def\as     {\ifmmode {\rlap.}$\,$''$\,$\! \else ${\rlap.}$\,$''$\,$\!$\fi}
\def\decsec  {\ifmmode {\rlap.}$\,$^{s}$\,$\! \else ${\rlap.}$\,$^{s}$\,$\!$\fi}\def\decs  {\ifmmode {\rlap.}$\,$^{s}$\,$\! \else ${\rlap.}$\,$^{s}$\,$\!$\fi}

\def\kms    {\ifmmode{{\rm km~s}^{-1}}\else{km~s$^{-1}$}\fi}

\def\ccm    {cm$^{-3}$}

\def\Lsun   {$L_{\odot}$}

\def\Msun   {$M_{\odot}$}
\def\Mspy   {\ifmmode {M_{\odot} {\rm yr}^{-1}} \else $M_{\odot}$~yr$^{-1}$\fi}
\def\Mdot   {\ifmmode {\dot M} \else $\dot M$\fi}
\def\mhd    {\ifmmode {n_{{\rm H}_2}} \else $n_{{\rm H}_2}$\fi}
\def\mhcd   {\ifmmode {N_{{\rm H}_2}} \else $N_{{\rm H}_2}$\fi}

\def\vlsr   {$v_{\rm LSR}$}

\def\El      {\ifmmode{E_{\ell}}\else{$E_{\ell}$}\fi}
\def\beam    {\ifmmode{\theta_{\rm B}}\else{$\theta_{\rm B}$}\fi}
\def\mjyb   {\ifmmode {{\rm mJy~beam}^{-1}} \else{mJy~beam$^{-1}$}\fi}
\def\Trot   {\ifmmode{T_{\rm rot}}\else$T_{\rm rot}$\fi}    
    
\def\Teff   {\ifmmode{T_{\rm eff}}\else$T_{\rm eff}$\fi}

\def\ITRS   {\ifmmode{\smallint {\rm T}_{R}^{*}dv}\else{$\smallint 
{\rm T}_{R}^{*}dv$}\fi}
\def\ITRS   {\ifmmode{\smallint {\rm T}_{R}^{*}dv}\else{$\smallint 
{\rm T}_{R}^{*}dv$}\fi}
\def\ITAS   {\ifmmode{\smallint {\rm T}_{A}^{*}dv}\else{$\smallint 
{\rm T}_{A}^{*}dv$}\fi}

\def\hh         {H$_2$}

\def\hzo        {H$_2$O}

\def\meth   {CH$_3$OH}

\def\ref#1  {\noindent \hangindent=24.0pt \hangafter=1 {#1} \par}
\def\ref#1  {\noindent \hangindent=18.0pt \hangafter=1 {#1} \par}
\def\lefttitle#1  {\noindent \hangindent=18.0pt \hangafter=1 {#1} \par}
\def\vol#1  {{\bf {#1}{\rm,}\ }}
\font\tenssb=cmssbx10
\font\tenbf=cmbx10
\font\sevenbf=cmbx8
\font\fivebf=cmbx6
\def\unetdemi    {\smallskipamount=6pt plus2pt minus2pt
                  \medskipamount=12pt plus4pt minus4pt
                  \bigskipamount=24pt plus8pt minus8pt
                  \normalbaselineskip=16pt plus0pt minus0pt
                  \normallineskip=2pt
                  \normallineskiplimit=0pt
                  \jot=6pt
                  {\def\smallskip {\vskip\smallskipamount}}
                  {\def\medskip   {\vskip\medskipamount}}
                  {\def\bigskip   {\vskip\bigskipamount}}
                  {\setbox\strutbox=\hbox{\vrule 
                    height17.0pt depth7.0pt width 0pt}}
                  \parskip 12.0pt
                  \normalbaselines}
\def\smallerspace {\smallskipamount=3pt plus0pt minus0pt
                  \medskipamount=6pt plus0pt minus0pt
                  \bigskipamount=10.5pt plus0pt minus0pt
                  \normalbaselineskip=10.5pt plus0pt minus0pt
                  \normallineskip=1pt
                  \normallineskiplimit=0pt
                  \jot=3pt
                  {\def\smallskip {\vskip\smallskipamount}}
                  {\def\medskip   {\vskip\medskipamount}}
                  {\def\bigskip   {\vskip\bigskipamount}}
                  {\setbox\strutbox=\hbox{\vrule 
                    height8.5pt depth3.5pt width 0pt}}
                  \parskip 0pt
                  \normalbaselines}
\def\memospace    {\smallskipamount=4pt plus1pt minus1pt
                  \medskipamount=6pt plus2pt minus2pt
                  \bigskipamount=14pt plus6pt minus6pt
                  \normalbaselineskip=14pt plus0pt minus0pt
                  \normallineskip=1pt
                  \normallineskiplimit=0pt
                  \jot=4pt
                  {\def\smallskip {\vskip\smallskipamount}}
                  {\def\medskip   {\vskip\medskipamount}}
                  {\def\bigskip   {\vskip\bigskipamount}}
                  {\setbox\strutbox=\hbox{\vrule 
                    height17.0pt depth7.0pt width 0pt}}
                  \parskip 2.0pt
                  \normalbaselines}
\def\memowidespace    {\smallskipamount=5pt plus1pt minus1pt
                  \medskipamount=7.5pt plus2pt minus2pt
                  \bigskipamount=17.5pt plus6pt minus6pt
                  \normalbaselineskip=17.0pt plus0pt minus0pt
                  \normallineskip=1.25pt
                  \normallineskiplimit=0pt
                  \jot=5pt
                  {\def\smallskip {\vskip\smallskipamount}}
                  {\def\medskip   {\vskip\medskipamount}}
                  {\def\bigskip   {\vskip\bigskipamount}}
                  {\setbox\strutbox=\hbox{\vrule 
                    height21.25pt depth8.75pt width 0pt}}
                  \parskip 2.5pt
                  \normalbaselines}
\begin{document}

\title{Very compact radio emission from high-mass protostars} 
\subtitle{I. CRL 2136: Continuum and water maser observations}
\author{K. M. Menten \&\ F. F. S. van der Tak}
\institute{Max-Planck-Institut f\"ur Radioastronomie,
Auf dem H\"ugel 69, D-53121 Bonn, Germany} 
%\\
%\email{kmenten, vdtak@mpifr-bonn.mpg.de}}
\offprints{K. M. Menten}
\date{Received / Accepted}
\titlerunning{Radio continuum and H$_2$O masers
toward CRL 2136}
\authorrunning{Menten \&\ van der Tak}

\abstract{
We report 5--43~GHz radio observations of the CRL 2136 region
at $0\as6$ -- $6''$ resolution. We
detect weak (mJy intensity) radio emission from the deeply embedded
high-mass
protostar IRS 1, which has an optically thick spectrum up to
frequencies of 22 GHz, flattening at higher frequencies, which might
be explained by emission from a jet. Water maser mapping shows that
the strong emission observed redshifted relative to the systemic
velocity is spatially coincident with the optically thick continuum
emission.
%which hides any blueshifted \hzo\ masers.
The \hzo\ maser emission from this object (and others we know of)
seems to have a different origin than most of these masers, which are
frequently tracing bipolar high-velocity outflows. Instead, the CRL 2136
\hzo\ emission arises in  the close
circumstellar environment of the protostar (within 1000 AU). We speculate 
that most of it is excited in
the hot, dense infalling gas after the accretion shock, although
this cannot explain {\it all} the \hzo\ emission. An accretion
shock nature for the continuum emission seems unlikely.

}

\maketitle

 \section{Introduction}
 For the earliest, deeply embedded phases of high-mass star formation,
 the distribution and kinematics of material on $\simless 1000$~AU
 scales is poorly known, due to the large ($\simgreat$ 1~kpc) distances
 involved, and the lack of tracers at optical and near-infrared
 wavelengths.
 The first systematic description of this phase was presented in a
 classic paper by Willner et al. (1982), which discussed 2--13 $\mu$m
 spectra of a sample of 19 compact infrared sources associated with
 molecular clouds already coined as ``protostars''. Observations of
 these sources by Mitchell et al.  (1990) indicated large column
 densities of CO, spread over multiple temperature and velocity
 components. In addition, Mitchell et al. (1991) and others discovered
 strong outflows with velocities up to 70 \kms\ in CO and/or \hzo\
 maser emission and up to 200 \kms\ in infrared CO absorption.

 Observations with the Infrared Space Observatory (ISO)
 Short-Wavelength Spectrometer (SWS, see, e.g., van Dishoeck et al. 1998) have
 refined our view on the structure and composition of the Willner et
 al.\ sources.  The ISO data suggest an evolutionary sequence starting
 with cold, ice-rich objects such as W33A and NGC 7538 IRS9, and going
 toward warm sources like GL 2591 where gas/solid molecular abundance
 ratios are $\gg$1.  Submillimeter maps (van der Tak et al.\ 2000b)
 indicate large masses of these envelopes, and a relation between
 temperature and the ratio of envelope mass to stellar mass.
 Additional data on this evolutionary sequence comes from submillimeter
 spectroscopy (van der Tak et al.\ 2000a, 2003), although the lines are
 more than an order of magnitude weaker than in the Orion ``Hot Core'',
 the prime example of its class, which is more evolved and less
 distant.

 Despite this progress, not much is known yet about the small-scale
 structure and kinematics of embedded high-mass protostars.  How do the
 observed outflows start? How do they interact with their environment
 on $< 1000$ AU scales? Subarcsecond resolution observations are
 necessary to shed light on these and other questions, which, at
 (sub)mm wavelengths, will begin to be addressable with the
 Submillimeter Array (Moran 1998), but for the high brightness sensitivities needed
 will  have to await the Atacama Large Millimeter
 Array 
(ALMA\footnote{http://www.eso.org/projects/alma/ or http://www.alma.nrao.edu/}).

 Centimeter-wavelength radio emission penetrates dust and can {\it now}
 be studied at these interesting resolutions with instruments such as
 the Very Large Array (VLA).  Medium-sensitivity (few mJy level) VLA
 surveys of many high-mass star-forming regions were made in the last 15
 years at high (sub-arcsecond) resolution (e.g., Wood \&\ Churchwell
 1989).  Subsequent (sub)millimeter-wavelength molecular line
 observations have established that in the ultracompact HII region
 (UCHII) region phase young high-mass stars are still surrounded by
 massive, dense, and hot molecular cores (Garay \&\ Lizano 1999).

 However, also found were a number of submillimeter sources undetected
 at cm-wavelengths at the sensitivity levels of the mentioned surveys,
 which, however from their IRAS colours, derived temperatures,
 densities, and dimensions were virtually indistinguishable from those
 associated with UCHII regions (e.g., Molinari et al. 1996, 1998, 2000;
 Sridharan et al. 2002; Beuther et al. 2002a) .  Sensitive VLA
  observations led to the detection of weak radio emission in some of
 these sources, which, in the case of the Turner-Welch object near the
 UCHII region W3(OH), surprisingly, turned out to be of non-thermal
 nature and exhibits a jet-like shape (Reid et al. 1995; Wilner et
 al. 1999).  In other cases, weak, up to (at least) 7 mm wavelength
 optically thick, ''hypercompact'' HII regions were found (e.g.,
 Tieftrunk et al. 1997; Churchwell 2002).  In the Orion-KL region, one such source (''I'') was
 found to be jet-like and have a thermal spectrum (see Menten \&\ Reid
 1995 and below).

 A basic motivation for searching compact radio continuum emission
 within high-mass protostellar cores is to precisely locate the
 position of the exciting sources, a critical requirement nowadays, as
 adaptive optics techniques deliver infrared observations with
 resolutions similar to the those of the interferometric radio data.
 Of equal or even more importance is the fact that the sheer existence
 of the radio continuum emission and its observed spectrum constrains
 theoretical models.

 A forthcoming paper (Van der Tak \& Menten, in prep.) summarizes
 existing radio data of high-mass protostars and presents new
 observations of such and similar sources.

 In this paper we consider the case of CRL 2136\footnote{Also known as
 AFGL 2136 or GL 2136; $l,b$= $17\decdeg639,+0\decdeg16$}
 evolution-wise an intermediate case between W33A and GL 2591.
 Multi-wavelength near-infrared (NIR) imaging by Kastner et al. (1992) 
 revealed a triple source structure, surrounded by
 nebulosity, which they whimsically named the ``Juggler Nebula''.
 Their IR-polarimetry led Kastner et al. to suggest that a deeply
 embedded source in the westernmost part of the triple structure, IRS
 1, was the dominating energy source, providing $5\times10^4$\Lsun\ to the region, for which they derive a kinematic
 distance of 2 kpc.

 % Compared 
 % to other sources in the Willner et al.  sample, foremost Orion IRc 2, this
 % one and other sources 
 % (such as W33 A or NGC 7538-IRS9) have  garnered comparatively little attention
 % from (sub)millimeter astronomers, although they are a cornucopia for infrared 
 % spectroscopists who profit from the strong background emission
 % to detect a large inventory of solid-state features.\footnote{Strictly 
 % speaking, this is true for W33A and NGC 7538-IRS9 and some other sources, 
 % but few searches 
 % for  solid-state features have been reported for CRL 2136, although
 % Kastner \&\ Weintraub 1996 report the detection of water ice.}
 % This is so because their (sub)millimeter
 % hot core molecular line emissions
 % are more than an order of magnitude weaker that those of the
 % prime  examples of the class.
 % Nevertheless, modern receiver technology allows observation of these
 % sources with weaker line emission and CRL 2136 was included in the source list of the
 % two studies conducted by van der Tak et al. (2000a,b), which, certainly not
 % by coincidence, consists almost entirely of Willner et al. sources.

 This paper reports successful multi-radio wavelength searches for weak
 continuum emission from CRL 2136. At our highest observing frequency
 (43.3 GHz) we resolve the emission. We also present maps of the
 unusually compact water maser emission distribution associated with
 the source.

 In Sect. 2
 %\ref{cha-data}  
 we describe the reduction of
 archival 
 %and newly acquired  
 VLA continuum and 22.2 GHz \hzo\
 maser line data of 
 CRL 2136 and present the results. 
 In Sect. 3 
 %\ref{disc}
 we  discuss these results in the context of other
 phenomena found in the region in question. We also claim that the 
 \hzo\ masers in this source belong to a class up to now
 not recognized, that  is excited in the innermost circumstellar
 regions rather, as most water masers, in outflows further out.

 \section{VLA observations, data reduction and results\label{cha-data}}
 \subsection{Archival continuum and water data}
 The archival CRL 2136 data discussed here were retrieved from the
 NRAO\footnote{The National Radio Astronomy Observatory (NRAO) is
 operated by Associated Universities, Inc., under a cooperative
 agreement with the National Science Foundation.}  Very Large Array
 (VLA) archival database (project name: AK 297, observing date: 1992
 May 19, when the VLA was in its $C$-configuration).  Three continuum
 $uv$-databases were obtained. Each had data taken with 2 intermediate
 frequency (IF) bands of width $2\times50$ MHz each centered $\pm 25$
 MHz of 4.8601, 8.4399, and 14.9399 GHz. (These band are termed {\it
 C-, X-,} and {\it U-}band  %C-, X-, and U-band 
in radio astronomy lingo).  In addition, a 127-channel
 spectral line database was used, containing a 15-minute duration
 snapshot of the \hzo\ maser line at 22.23508 GHz toward CRL 2136
 (which is in ``K''-band). Each of the channels was of 12.2 kHz
 width, corresponding to 0.165 \kms\ and the total velocity coverage
 was 20.8 \kms, centered at an LSR velocity of 26.3 \kms\ and, thus,
 covered $-6.9$ to $+13.9$ \kms\ around the systemic velocity, which is
 22.8 \kms\ (van der Tak et al. 2000b).

 The data were edited, calibrated, and imaged in the ``usual'' way with
 NRAO's Astronomical Image Processing System (AIPS). In the case of the
 spectral-line data, the ``channel 0'' database, comprising the inner 75
 \%\ of the passband, was used for this. Absolute calibration was
 obtained from observations of 3C286 using the fluxes interpolated from
 the values given by Baars et al. (1977). NRAO 530 was the phase
 calibrator.  Unfortunately, no 22.2 GHz data exist for 3C286.  To
 achieve absolute calibration for the K-band data we determined NRAO
 530's flux density at that 22.2 GHz by extrapolation using the
 spectral index, $\alpha_{\rm XU},$ determined from its X- and U-band
 flux densities. The error in the absolute calibration should be within
 10\%.  %
 %Actually NRAO 530's flux in  c.u. was 1.1 Jy, which we multiplied by 6
 % S_C= 7.5
 % S_X= 5.1
 % S_U=6.5 Jy (from PRTAB of SU table)
 % Weird spectrum!

 Restoring beam major and minor axes and position angles (PAs,
 east of north) were (7\as7,4\as5,18\deg), (4\as3,2\as5), (2\as4,1\as5,6\deg)
 for the C-, X-, and U-band maps, respectively. 
 %and 
 %and (1\as6,1\as1,-22\deg) and (0\as56,0\as041,7\deg) for the Q-band 
 %C- and A-array maps, respectively.

 \subsection{Effelsberg water observations}
 A spectrum of the \hzo\ maser line was taken in position-switching
 mode with the MPIfR 100 m telescope near
 Effelsberg, Germany. A K-band HFET receiver was used and the
 spectrometer was an autocorrelator.

 \subsection{New VLA continuum data}
 New VLA data were taken on two dates: On 2001 September 09 in
 C-configuration and on 2002 March 23 in A-configuration. This time
 observations were made in the highest (''Q'') VLA frequency band. On
 both dates, data were taken with 2 intermediate frequency (IF) bands of
 width $2\times50$ MHz each centered $\pm 25$ MHz of 43.3399
 GHz. Absolute and phase calibration and data processing was performed
 as described above, with the difference that short (duration $\sim 70$
 sec) scans of CRL 2136 were alternated with 10 sec duration scans of
 the nearby calibrator 18296$-$10374, with a 20 sec ''dead time'' in
 between (needed for slewing). As proven {\it a posteriori} by the
 quality of the resulting images, this fast ''switching'' 
 provided for near-perfect calibration. For absolute calibration 3C286
 was observed.  Q-band observations are more affected by variations
 in the weather conditions and gain variations with telescope elevation
 than lower frequency observations..  Therefore, we estimate a 30\%\
 uncertainty for our Q-band flux densities.

 Restoring beams were  (0\as61,0\as43,$-2$\deg)
 and (0\as057,0\as041,7\deg) for the Q-band 
 C- and A-array maps, respectively. 

 \subsection{Continuum results}

 \begin{table*}[tb]
 \caption{Radio and infrared  emission from CRL 2136}
 \label{tab:sources}
 \begin{tabular}{llllllllll}
 \hline \hline
 Source &
 $\alpha_{2000}$ & 
 $\delta_{2000}$& 
 $\sigma (\alpha,\delta)$& 
 $S_{\rm p,4.9}$ & 
 $S_{\rm i,4.9}$ & 
 $S_{\rm p,8.4}$ & 
 $S_{\rm i,8.4}$ & 
 $S_{\rm p,14.9}$ &
 $S_{\rm i,14.9}$\\
 &18\h22\m&-13\deg&
 (arcsec) &
 (mJy $b^{-1}$) &
 (mJy)&
 (mJy $b^{-1}$)&
 (mJy)&
 (mJy $b^{-1}$) &
 (mJy) \\
 \noalign{\smallskip}
 \hline
 \noalign{\smallskip}
  RS 1 & 17\decs22  &27$'$ 41\decas4  & 0.3  & 3.1(0.2) & 4.5(0.5) & $<0.14$ & -- & $<0.42$ & -- \\
  RS 2 & 22.11 & 33 18.5 & 0.4  & 1.9(0.2) & 2.9(0.5) & $<0.14$ & -- & $<0.42$ & -- \\
  RS 3 & 26.10 & 30 09.7  & 0.3  & $<0.48$ & -- &0.63(0.08) & 0.69(0.16) & $<0.5$ & -- \\
 \meth & 26.3 & 30 06& 10 \\
  RS 4 & 26.37 & 30 11.9  & 0.3  & $<0.48$ & -- & 0.56 (0.08) & 0.56 (0.08) & 0.99(0.13) & 1.30(0.27) \\
 H$_2$O& 26.38 & 30 11.8 & 0.1 \\
 IRS 1 & 26.5 &  30 12 & 2  \\
 RS5   & 39.209 & 29 56.36 & 0.2 & 9.0(0.2) & 9.1(0.3) & $<0.4$ & -- & $<0.5$ & -- \\
 \noalign{\smallskip}
 \hline
 \noalign{\smallskip}
 \end{tabular}

 $^a$ $S_{{\rm p},\nu}$ and $S_{{\rm i},\nu}$ are the peak and integrated flux
 densities, respectively, and their {\it formal} errors determined with
 the AIPS task JMFIT.  The {\it absolute} flux density calibration
 should have uncertainties of less than 10\%\ at C, X, and U-band and
 $\simless 20$\%\ at K- and $\simless 30$\%\ at Q-band.  
 Upper limits are 3 times the
 $1\sigma$ rms noise. Radio position errors are formal errors from
 JMFIT with the estimated absolute position error of 0.1 arcsec
 quadratically added. The error in the average \hzo\ maser position is
 completely dominated by the latter. This ``average'' \hzo\ position is
 the variance-weighted mean position of the Gaussian fit results to the
 velocity channels with strong ($> 10$ Jy) emission, i.e., from 26.5 to
 27.9 \kms.  RS 4 is also detected at 43 and 86 GHz at a position
 consistent with that given above (see Table~2). See reference in text for the
 \meth\ maser position.

 \end{table*}

 At 4.9, 8.4, and 14.9 GHz we produced large maps of size
 $2048''\times2048''$ around the phase center position of
 $(\alpha,\delta)_{{\rm J}2000}$ $= 18\h22\m26\decs482, -13\deg30'13\decas1$.  
 At 43.3 GHz only a small map of extent (6\as4,6\as4) was
 made.  Table~1 lists the sources detected (using the
 multiple-peak-finding AIPS task SAD) with more than 5 times the
 $1\sigma$ rms noise levels of 0.16, 0.046, and 0.14 mJy~beam$^{-1}$ in
 the C-, X-, and U-band maps. The flux densities are corrected for
 primary beam response.  The errors in Table~1 are
 statistical errors delivered by SAD added quadratically to the
 absolute errors of $0\as1$ determined as follows. To obtain a
 ``realistic'' estimate of the absolute position errors remaining after
 phase calibration, we ``calibrated'' the NRAO 530 data ``with'' the
 3C286 data. The resulting position differed from 3C286's nominal
 position by $(\theta_{\rm x},\theta_{y}$) = $(+0\decas08,-0\decas66)$,
 $(+2\decas0,+0\decas62)$, and $(0\decas54,-0\decas24)$ at C-, X-, and
 U-band, respectively. Since the arc between NRAO 530 and CRL 2136 is
 more than 50 times greater than that between NRAO 530 and CRL 2136, we
 feel safe to assume that the errors in the positions in
 Table~1 are smaller that 0.1 arcsec.
 %A total of 5 radio continuum sources are detected and listed in Table 1
 %\ref{tab:sources}. 
 Rather than indicating that radio sources 1, 2, 3, and 5 in Table 1
 are marginally resolved, the slightly higher numerical values of the
 integrated flux densities than the peak values are most likely caused
 by residual phase errors; self calibration was not possible due to the
 weakness of the emissions.  Three radio sources (RS 1, 2, and 5) are
 detected at 4.9 GHz only and one (RS 3) at 8.4 GHz only.  As shown in
 Fig. 1, only RS 3 and 4 are in the immediate vicinity of IRS 1.

 \begin{figure}
 \begin{center}
 \includegraphics[height=11cm]{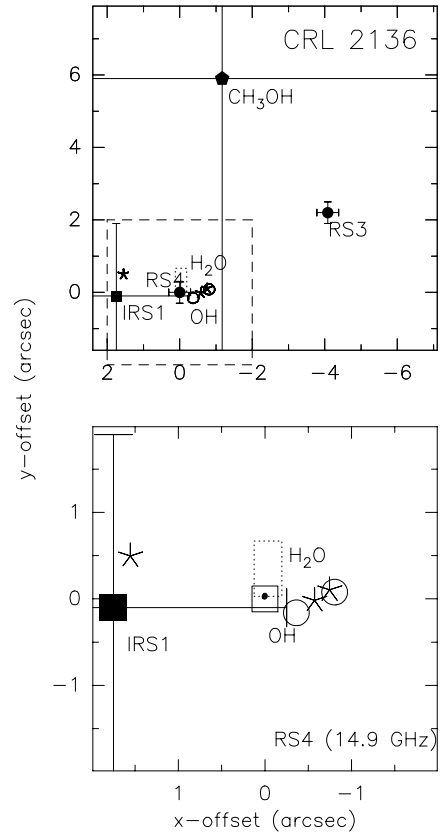}
 \caption{Objects in the CRL 2136 region: The {\it upper panel} shows
 the positions of radio sources 3 and 4 {\it (filled circles)},
 infrared source 1 {\it (filled square)}, and the class II \meth\ maser
 {\it (filled pentagon)}.  The three little stars mark the positions
 from which of 1665 MHz RCP maser emission arises and the circles mark
 LCP emission.  For the OH emission, the symbol sizes reflect the $0\as3$
 position uncertainties.  The (at this scale) unresolved dot at
 the (0,0) position represents the Q-band map of RS 4, an enlarged
 image of which is shown in Fig. 2.  The square around it indicates its
 $0\as3$ position uncertainty.  The {\it dotted rectangle} outlines the
 region containing \hzo\ maser emission shown in detail in Fig. 4 and
 the {\it dashed rectangle} outlines an enlarged region shown in the
 {\it lower panel}.}
 %, in which the contours represent the
 %R20414.9 GHz radio emission. Contour values are 3, 5, and 7 times the
 %$1\sigma$ rms noise of 0.14 mJy${\rm beam}^{-1}$. Other symbols have
 %the same meaning as in the {\it upper panel}.}
  \label{fig1}
 \end{center}
 \end{figure}

 \subsubsection{Radio source 4}
 Using JMFIT, we determine Q-band peak ($S_{\rm p}$) and integrated
 ($S_{\rm i}$) intensities of 4.49 (0.22) \mjyb\ and 4.20 (0.36) mJy,
 respectively, for the C-array data and 1.43(0.10) \mjyb\ and
 1.78(0.21) mJy for the A-array data.  The discrepancy between the A-
 and C-array fluxes is larger than our assumed error margin and is
 possibly due to source variability.
 We produced a source model, derived from a Gaussian fit to our
 (lower resolution) C-band data and introduced it in the A-array
 $u,v$-database, imaged it and made a Gaussian fit to the model source
 flux distribution in the A-array map. We retrieved {\it all} the  input flux.
 We therefore are certain that  we are not ``resolving out'' any extended 
 structure and have great confidence about the quality of our phase calibration.

 JMFIT statistics yields the source properties listed in Table 2 with %(large) 
 formal errors. The dimensions in the table %(plus their) errors
 should probably taken as an upper limit on the actual source size.
 %% size of $0\as041\times0\as033$ and a minimum size of
 %% $0\as0\times0\as0$, with a nominal value of
 %% $0\as029\times0\as003$. Maximum and minimum values of the PA are
 %% 148\deg\ and 73 \deg, respectively (nominal value 116\deg).  
 Fig.~2 illustrates these results.  The shape (elongation) and size as
 well as its radio luminosity and spectral index make RS 4, which is of
 central interest to this paper, strikingly similar to source I in the
 Orion-KL region (see Table 2 and discussion below).

 \begin{figure}
 \begin{center}
 \includegraphics[height=7cm]{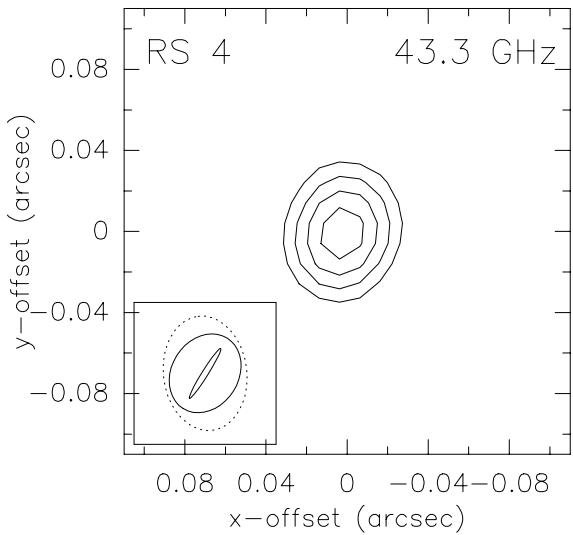}
 \caption{\footnotesize 
 Radio source 4 at 43.3 GHz: The contours represent  5, 7, 9, and 11 
 times the 2.7 \mjyb\ rms noise in our 43.3 GHz map. The {\it dotted ellipse}
 in the {\it lower left} corner represents the FWHM size of the 
 Gaussian restoring beam, 
 while the {\it full line} ellipses represent the maximum and nominal 
 FWHM source size from JMFIT (see discussion in text).
 Angular offsets are relative to the position given in Table 1.}
 \label{fig:fig2}
 \end{center} 
 \end{figure}

 %We calculate an $\alpha_{\rm UW}$ of $2.2^{+0.4}_{-0.5}$, assuming a 50\%\
 %uncertainty in the W-band value.
 %It is this source which is of central interest to this paper.

 \subsection{Water results}
 In the water line, a $25\decas6\times25\decas6$-sized map was made of
 the channel with the strongest emission (at 27.1 \kms).  After several
 iterations of self calibration a 390 Jy strong feature was obtained.
 The phase and amplitude corrections were copied to the other velocity
 channels. 

 \begin{figure}
 \begin{center}
 \includegraphics[height=13cm]{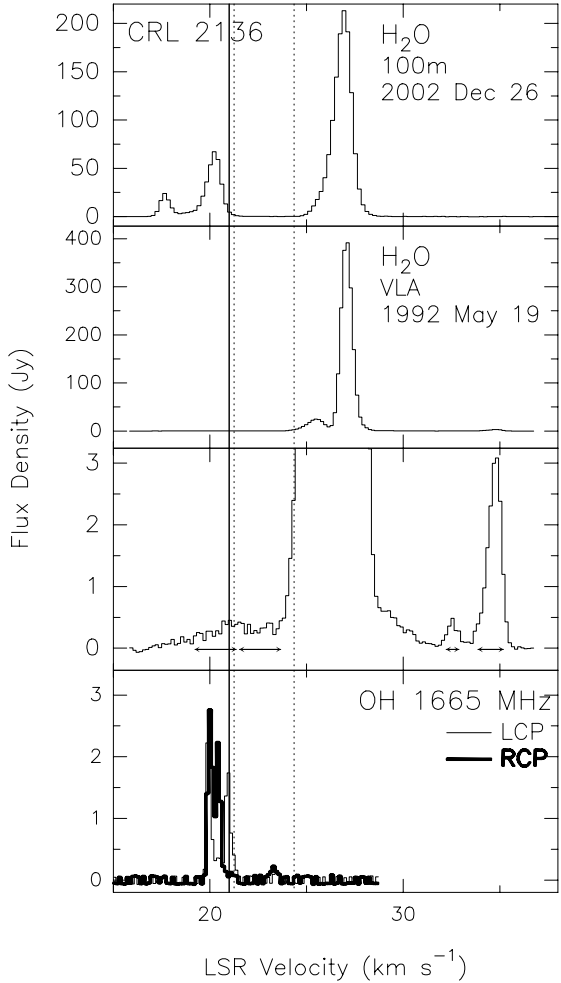}
 \caption{\footnotesize
 {\it Top panel:} \hzo\ spectrum taken at the Effelsberg 100 m 
 telescope on 2002 December 26.  {\it Second panel from  top:}
 VLA \hzo\ spectrum made from our 1992 May 19 data at the pixel 
 with maximum emission.
 In the {\it third panel from top} the intensity axis is expanded to
 show the extremely weak emission at the lowest and highest
 velocities. The arrows show the velocity ranges over which maps
 of integrated emission, marked by the rectangles in Fig. 4
 were produced. The {\it lower panel} shows the RCP and LCP emission
 of the OH 1665 MHz line published by Argon et al. 2000.
 For details of how the spectrum was formed see this 
 reference.
 The dotted vertical lines mark the FWHM linewidths of  various
 molecular species mapped by van der Tak. et al (2000b), who 
 determine a centroid LSR velocity of 22.8(0.1) \kms.  The 
 solid line marks the velocity of the single, narrow 6.7 GHz 
 \meth\ maser feature (Caswell et al. 1995).} \label{fig:fig3}
 \end{center}
 \end{figure}

 A spectrum at the pixel with maximum emission was made and is shown in
 Fig.  3.  The peak emission is comparable in strength to the 330 Jy
 Valdettaro et al. (2001) report between 23 and 31 \kms; the weak
 emission we observe outside that range is below their sensitivity
 limit.  Their velocity of peak emission, $v_{\rm peak} = 27.1$ \kms,
 coincides exactly with ours, which is remarkable, since their
 observation was made on 2000 January 18, almost 8 years after the VLA
 data discussed here were taken. On 1991 January 31, Kastner et
 al. (1992) also detect a single feature (of 46 Jy flux density), 
 however at \vlsr = $26.3 \pm
 0.1$ \kms.
 %If they {\it really} used 22235 MHz as frequency for the \hzo\
 %line, instead of 22235.08 MHz, their line would appear at an 
 %1.1 \kms\ too low velocity. Correcting for this brings their
 %velocity to $27.4 \pm\ 0.1$ \kms, which is practically
 %identical to the velocities measured on the other dates.
 The Effelsberg spectrum has its peak emission at 27.0 \kms.  In some
 sources \hzo\ maser spectra vary significantly on timescales as short
 as days and it is worth noting that the ``stability'' of a \hzo\ maser
 spectrum observed toward CRL 2136 appears unique and has to be
 considered when modeling the emission.

 While in the VLA spectrum, all of the strong emission is redshifted
 relative to the systemic velocity, the Effelsberg spectrum shows
 redshifted (between 24.6 and 28.8 \kms) as well as moderately strong
 blueshifted emission (between 16.7 and 21.7 \kms) at our $5\sigma$
 noise level of 0.8 Jy and nothing between nor outside of these
 velocity intervals.

 \begin{figure}[p]
 \begin{center}
 \includegraphics[height=19cm]{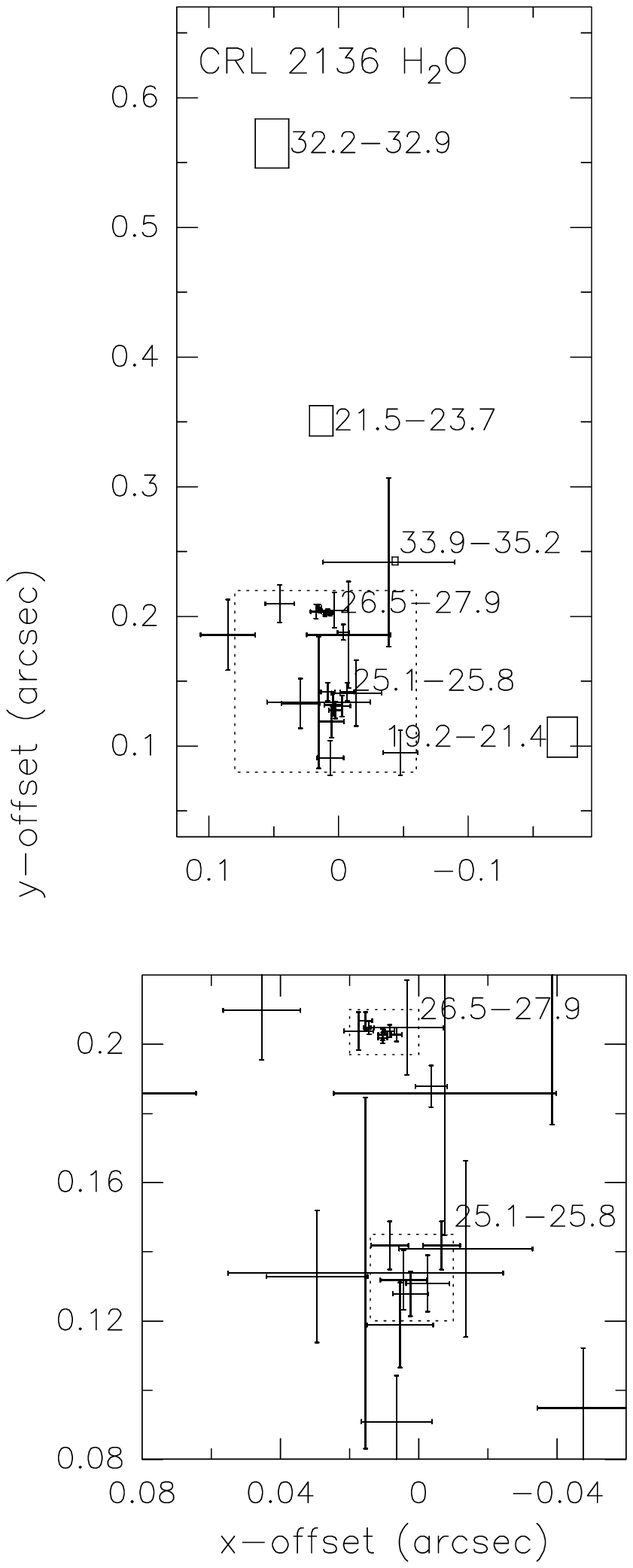}
 \caption{
 Area containing the \hzo\ maser emission, indicated
 by the {\it dashed rectangle} in Fig. 1.
 The  {\it error bars} in the {\it upper panel} 
 give the positions of \hzo\ masers determined from channel by
 channel fitting of the \hzo\ data cube. The centers of the {\it full line}
 rectangles are   the positions of the centroid positions of 
 maps integrated over the
 velocity ranges given (and indicated in Fig. 3); the sizes corresponds to the $1\sigma$
 errors of the fitted position. The {\it dotted rectangle}
 encloses the two regions of strongest emission, whose velocity
 ranges are given. This area is shown in more  detail
 in the {\it lower panel}, in which the emission in the {\it dotted rectangles}
 spreads over the indicated velocities. Position offsets are
 relative 
 %to the centroid of the emission in the 25.1 to 27.9
 %\kms\ velocity range, which is at a formal offset of
 %$(\theta_{\rm x},\theta_{\rm y}$ = $ +0.01,+0.15$ from, and thus
 to the position of RS 4 given in Table 1.
 Note that the relative registration of the \hzo\ and RS 4
 have an estimated error of 0.3 arcsec.}
 \label{fig:fig4}
 \end{center}
 \end{figure}

 In Fig. 4 we present the results of Gaussian position fitting (using
 the AIPS task JMFIT). Since the total spatial spread of the emission
 is all within a synthesized beamwidth, only a single Gaussian was
 fitted to each channel. Each cross presents the position and $1\sigma$
 error bars of the emission in one channel. The arrows underneath the
 spectrum (Fig. 3) mark parts which were averaged before mapping. The
 emission from these velocities falls in the same general region as the
 stronger emission mapped individually, i.e., the $0\as3\times0\as5$
 ($600\times1000$ AU) region around the variance-weighted mean position
 which is given in Table~1.  Based on our experience
 with the lower-frequency data described below, we estimate that the
 ``real'' {\it absolute} position uncertainty (rather than the formal
 fitting error) in each coordinate is of order $0\decas2$.  To check
 whether other emission was present in the \hzo\ channel maps, we used
 the aforementioned AIPS task SAD. We searched for any peaks with flux
 density above five times the $1\sigma$ rms noise level in each channel, which 
 was between 36 and 54 \mjyb\ for most channels except for $\pm 4$ channels
 around the channel with strongest emission (at 21.7 \kms) in which it
 had a maximum of 220 \mjyb, bearing witness to the excellent dynamic
 range in this (snapshot) observation.
 No emission was found except for that shown in Fig. 4, which covers
 an area of $\simless 10\%$ of our FWHM synthesized beam width.

 \section{Discussion}
 \label{disc}

 \subsection{OH and \meth\ maser emission from the CRL 2136 hot core}

 The molecular hot core around CRL 2136 is characterized by elevated
 temperatures. For example, van der Tak et al. (2000a) find a methanol
 rotation temperature of 143 K, which is a lower limit to the kinetic
 temperature of the \meth-emitting gas. However, the \meth\ abundance
 is only $9\times10^{-10}$ relative to \hh, 2 to 3 orders of magnitude
 below the solid-state abundances and the gas-phase values found in
 other hot cores of similar temperature and difficult to explain, since
 \meth\ evaporates off grain mantles for temperatures exceeding $\sim
 100$ K (Sandford \&\ Allamandola 1993). Possibly, a
 compact hot core with much higher \meth\ abundance exists, but,
 if it has a diameter of, say, $1''$ would have a much smaller
 filling factor in van der Tak et al.'s $14''$ beam and  would be 
 ``swamped'' by more extended, lower methanol abundance material.
 Such a hot core would need to have much higher temperature
 than 143 K, because that value should in this case represent an ``average'' 
 of the temperatures of  the extended and  hot core material.

 Some insight may come from \meth\ masers, which can be studied at
 milli-arcsecond resolution. As shown by Walsh et al. (2001) and
 Beuther et al. (2002b), class II \meth\ masers are excellent tracers of
 deeply embedded massive and intermediate-mass stars, as well as of
 ultracompact HII regions. The 6.7 GHz class II \meth\ maser emission
 toward CRL 2136 found by Macleod et al.  (1992) in 1991
 October/November consisted of a single, narrow ($\Delta v < 1$ \kms),
 25 Jy strong feature at the position listed in Table 1 and is, given
 its $10''$ error in both coordinates coincident with our radio sources
 3 and 4, although an identification with RS 4 seems more likely.
 Caswell et al. (1995) reobserved a virtually identical spectrum in
 1992/1993.  A more accurate position determination and high resolution
 mapping of the CRL 2136 \meth\ maser, e.g. with Australia Telecope
 National Facility Compact Array \footnote{http://www.atnf.csiro.au/},
 seem highly desirable.

 Weak (1--2 Jy) OH maser emission was observed by Cohen et al.\ (1988)
 between $\approx 19$ and 23 \kms\ in the 1665 and 1667 main hyperfine
 transitions and the former line was mapped using the VLA by Argon et
 al.\ (2000).  The latter authors' positions are shown 
%with our 14.9 GHz continuum map of RS 4 
in the lower panel of Fig. 1. It is obvious
 that the OH, and by implication the class II \meth\ masers probe the
 hot, dense environment in the immediate vicinity of RS 4. Other proof
 of this is the good agreement of the velocity spread of the lines from
 various species studied by van der Tak et al.\ (2000b).

 The position of one 1665 GHz LCP feature (at 20.92 \kms) agrees to
 within 3 times the $1\sigma$ relative position uncertainly with that
 of an RCP feature at 20.42 \kms, indicating Zeeman-spitting. The
 derived B-field strength is 1 mG, which is a few times smaller than
 ``typical'' values one finds in interstellar OH maser regions.
 % 0.59km/s/mG ZEEMAN 1665

 To summarize the OH and \meth\ maser data: The maser velocities
 suggest that they arise from the hot core surrounding IRS 1, but they
 are in a region further away (1000-- 4000 AU) from the heating source
 (IRS 1), in which temperatures ($\sim 150$ K) and densities ($\sim
 10^7$ \ccm) (see Menten 1997) are conducive for their excitation.

 \subsection{Radio source 4 and its environment}

 \def\pz {\phantom{0}}
 \begin{table*}[tb]
 \begin{center}
 \caption{Comparison: CRL 2136-RS4/Orion-KL (I)}
 \label{tab:comp}
 %\rotate
 %\tiny
 \begin{tabular}{lll}
 \hline \hline
  & CRL 2136-RS 4    & Orion-KL (I) \\
 \noalign{\smallskip}
 \hline
 \noalign{\smallskip}
 $S_{\rm i,4.9}$ (mJy)   &\pz$<0.48$\pz\pz\pz &-- \\
 $S_{\rm i,8.4}$ (mJy)  &\pz\pz$0.56\pm0.08$  &\pz\pz$ 1.1\pm0.2$ $^b$ \\
 $S_{\rm i,14.9}$ (mJy)  &\pz\pz$1.3\pm0.3$   &\pz\pz$1.6\pm0.4$ $^c$ \\
 $S_{\rm i,43.2}$ (mJy)  &\pz\pz$4.2\pm0.3$   &   \pz$16\pm2$    $^d$ \\
 $S_{\rm i,86}$ (mJy)    &\pz$61\pm18$ $^e$   &   \pz$34\pm5$ $^f$\\ 
 $\theta_{\rm a}$ (mas)  &\pz$29^{+12}_{-29}$  &   \pz$96\pm10$ $^d$\\
 $\theta_{\rm b}$ (mas)  &\pz\pz$3^{+30}_{-3}$ &\pz$65\pm9$ $^d$\\
 major axis a (AU)        &\pz58 &  \pz43 \\
 minor axis b (AU)        &\pz\pz6 & \pz 29 \\
 position angle (E of N)& $116^{+32}_{-43}$& $146\pm12$ $^d$ \\
 \noalign{\smallskip}
 \hline
     \end{tabular}
   \end{center}
 \medskip

 $^a$ $S_{p,\nu}$ and $S_{i,\nu}$ are the peak and integrated flux
 densities, respectively, determined with the AIPS task JMFIT.  The
 upper limit is 3 times the $1\sigma$ rms noise.  Source sizes were
 also determined using JMFIT and are discussed in the text.

 $^b$ Menten \&\ Reid 1995

 $^c$ Felli et al. 1993

 $^d$ Menten \&\ Reid (in prep.)

 $^e$ van der Tak et al. 2000b

 $^f$ Plambeck et al. 1995

 \end{table*}

 Although the relation of the intensity of the weak radio continuum
 emission detected in some high-mass protostars to their overall
 luminosity is at present not understood at all, it is clear that it
 represents a signpost for the exact position of the object and should,
 by means of theoretical modeling, give clues to its evolutionary
 state.  Take the case of Orion-IRc2: Here, as shown by Menten \&\ Reid
 (1995), the weak radio emission from source ``I'' almost certainly
 marks the position of the exciting object in the region, as the
 excitation of its surrounding SiO masers requires extreme temperatures
 and densities.  Apart from the important signpost function, the
 detection of radio emission, which in the case of I is optically thick
 at least to a frequency of 43 GHz, can also put interesting
 constraints on the nature of the embedded protostar, as recently shown
 by Tan (2003), who proposes a jet model to explain I's radio emission.
 It should be noted that at the distance, $D$, of CRL 2136 (2 kpc), the
 detection of Orion source I ($D = 450$ pc) would require many hours of
 VLA time.

 With regard to its low radio luminosity, rising spectrum, and likely
 connection with a powerful infrared source, RS 4 is similar to source
 I in Orion-KL (Table 2 and Figure 5).  The latter source's position in
 the centroid of SiO maser emission, which require high temperatures
 ($\sim 1000$ K) and densities ($10^9$ \ccm) to be excited over the
 extent observed, makes it clear that it is the powering source in the
 region, providing the major portion of its luminosity of $\sim 10^5$
 \Lsun\footnote{No SiO maser was found toward CRL 2136 by Kastner et
 al. 1992 in the 43.2 GHz $v=1, J=1-0$ transition to an rms noise level
 of 1 Jy. SiO maser emission has only been been detected in 3
 star-forming regions, and in one of these, W51N (see Eisner et
 al. 2002) {\it only} in the $v=2, J=1-0$ transition.}  Also remarkable
 is that in Orion-KL the radio continuum/SiO emission is offset from
 the infrared source IRc 2, which represents reprocessed radiation,
 while the extinction toward the ``real'' source is so high to render
 it invisible even at the longest IR wavelengths accessible from the
 ground.

 One may ask whether similar geometrical circumstances also apply to
 CRL 2136 and other infrared protostars. At first sight, it certainly
 seems peculiar that some of the most massive compact dust emission
 sources are even detectable at near- and mid-IR wavelengths, and,
 moreover, shine as strong background sources for spectroscopy
 throughout that wavelength range, while others, such as the
 intermediate-mass protostar(s) W3(OH)-TW, are completely undetectable
 from the ground (Wyrowski et al. 1997, 1999; Stecklum et al. 2002).
 For the Willner et al.\ sources, the CO column densities in infrared
 absorption (pencil beam) and submm emission (15$''$ beam) differ by
 factors of 3 to 5, limiting the importance of geometry (van der Tak et
 al.\ 2000b).

 %% Certainly some of the Willner et al. sources, in particular W33 A,
 %% have been so intensely studied by IR-astronomers
 %% just {\it because} they are such easy
 %% (strong) targets. 
 %We defer this  discussion to \S \ref{others}
 %and go discussing our continuum and \hzo\ maser data.

 What is the nature of the observed radio continuum emission from CRL
 2136 (Fig. 5)?  We derive a lower limit to the brightness temperature
 of 2020 K at 43.3 GHz, much higher than the dust sublimation temperature.  
 RS 4's 8.4 to 14.9 GHz
 spectral index (SI), $\alpha_{\rm XU}$ is 1.5; inclusion of the 43.3
 GHz data point yields $\alpha_{\rm XQ}$ = 1.2 and we are, thus, in all
 likelihood seeing optically thick(ish) free-free emission up
 frequencies of 14.9 GHz, which becomes optically thinner at higher
 frequencies.

 \begin{figure}[tp]
 \begin{center}
 \includegraphics[width=8cm]{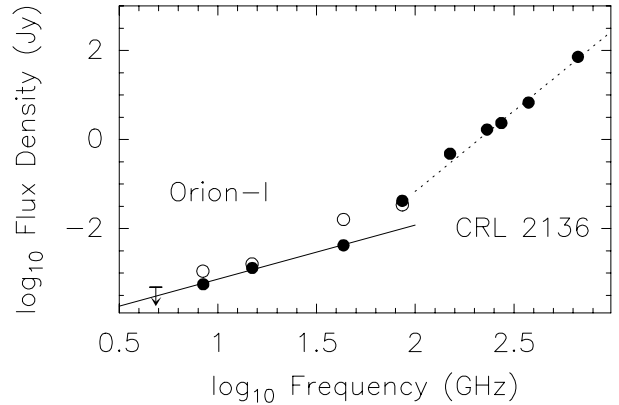}
 \caption{\footnotesize
 The radio-to-submillimeter-wavelength spectral
 energy distributions of CRL 2136 IRS 1/RS 4 ({\it full dots})
 and Orion source I ({\it open dots}).
 The {\it full line} represents a fit to our 4.9 to
 43.3 GHz data for RS 4, while the {\it dashed line} is a fit
 to the submillimeter data of Kastner et al. 1994. 
 Error bars of flux densities are smaller than or comparable to the
 symbol sizes.
 }
 \label{fig:fig5}
 \end{center}
 \end{figure}

 From the Kastner et al. (1994) 150 to 667 $\mu$m data we determine an
 SI, $\alpha_{\rm submm}$ of 3.3, consistent with optically thin dust
 emission (Fig. 5).  Extrapolation down to 86 GHz from the higher
 frequencies using this SI shows 95\%\ of the 86 GHz flux density of 42
 mJy can be accounted for by dust emission. Extrapolation of the
 cm-data (using $\alpha_{\rm XQ}$ = 1.2) yields 17 mJy, indicating a
 flattening of the radio spectrum.  As Table 2 and Fig. 5 demonstrate,
 RS 4 and I share a number of characteristics: Both have elongated,
 jet-like morphology, rising radio spectra that flatten above 30 GHz,
 and even comparable physical dimensions.

 Motivated by Orion source I's elongated morphology, Tan (2003) models
 this source as a double jet, powered by accretion onto a
 protostar. For accretion rates above $10^{-4}$ \Mspy\ he finds the
 flux densities at $\nu\lax$ 30 GHz to be insensitive to the accretion
 rate and to grow as $\nu^2$, while the spectrum flattens at higher
 frequencies and a $\nu^2$ dependence up to 86 GHz seems to require
 $\Mdot_{\rm in} > 5\times10^{-4}$ \Mspy, a very high value. Maybe this
 model can be taken as a starting point for further studies
 trying to explain the radio emission from IRS 1/RS 4. Differences
 between RS 4 and I are the former source's $\sim 10$ times higher
 lower frequency radio luminosity and the lower turnover frequency.
 Reynolds (1986), modeling collimated ionized stellar winds, finds
 that an SI of 1.2 requires recombination or acceleration in the flow.

 Neufeld \&\ Hollenbach (1996) have calculated the free-free emission
 emerging from an accretion shock for different protostellar masses,
 $M$, and infall rates $\Mdot_{\rm in}$.  Alas, even for the most
 extreme cases they considered, $M$ = 10 \Msun\ (solar masses) and
 $\Mdot_{\rm in} = 10^{-4}$ \Mspy, the 4.8 and 8.4 GHz flux densities
 they obtained, scaled to $D = 2$ kpc, are more than a factor of
 hundred lower than the flux densities we measure at these frequencies,
 making this mechanism extremely unlikely.

 \subsection{CRL 2136's compact water maser emission}

 The three most salient observational facts of the \hzo\ maser emission
 from RS 4 are: First, the strongest \hzo\ maser mission is redshifted
 relative to the systemic velocity of 22.8 \kms.  Second, the centroid
 position of the \hzo\ emission has a formal offset of $(\theta_{\rm
 x},\theta_{\rm y})$ = $(+0.01,+0.15)$ from, and thus is within the
 errors {\it coincident} with RS 4, and third, {\it all} of the \hzo\
 emission is within the small rectangle of size $0\as3\times0\as5$
 ($600\times1000$ AU) outlined in Fig. 1 and shown in detail in Fig. 4.
 It seems worth mentioning that the strongest ($S>10$ Jy) emission
 arises from two compact regions, separated in NS direction by $0\as07$
 (= 140 AU).

 Assuming, first, that the LSR velocity of the protostar is identical
 to that of its surrounding hot core, and, second, that either 
 of the compact strong emission regions is spatially coincident with
 the continuum source, the observed redshift of the strong \hzo\
 emission means that we are observing infall of the water-containing
 material onto the protostar.  The fact that the continuum emission is
 optically thick naturally explains the predominance of red-shifted
 \hzo\ emission.

 If the above were true, the case of CRL 2136 would be a
 convincing case of water maser emission emerging from {\it inflowing}
 circum-(proto)stellar gas. There is abundant evidence for interstellar
 maser emission associated with bipolar molecular {\it out}flows, both,
 from the frequently observed extreme velocity ranges (often $\pm$ tens
 of \kms\ around the systemic velocity or more) and from direct proper
 motion determinations (see, e.g., Reid \&\ Moran 1988).  For the
 (outflow) cases, excitation calculations place the water in the
 postshock regions of $J$-shocks, where densities are a few times
 $10^8$ -- $10^9$ \ccm, the temperature is $\sim 400$K, and the water
 abundance is enhanced in the postshock chemistry (Elitzur et al. 1989).

 Given the fact that \hzo\ masers are ``usually'' outflow tracers {\it par
 excellence} we speculate, hesitantly, but excitedly, that the redshifted
 \hzo\
 maser emission moving onto CRL 2136 is produced in the postshock gas
 behind the accretion shock. Neufeld \&\ Hollenbach (1996) consider
 accretion shock velocities of 30 \kms\ or higher, much smaller than the values
 indicated by our emission.

 Given that CRL 2136 drives a massive outflow (with 50 \Msun\ in the
 outflowing gas), mapped in the CO molecule by Kastner et al. (1994),
 it is curious that we do not see any water emission at all that is
 associated with the outflow, given the ubiquity of \hzo\ maser
 emission observed from outflows from stars of all masses\footnote{The
 poorly collimated CO outflow is along a position angle of $\sim
 135\deg$ over a $\sim 80''\times80''$ ($0.8\times0.8$ pc) area, with
 the blueshifted gas SE and the redshifted NE of IRS 1. There is little
 correspondence with our \hzo\ maser distribution (Fig. 4), other than
 that we also observe redshifted emission N of the blueshifted emission,
 although at 200 times smaller scale.}.  Most regions, placed at CRL
 2136's distance, would show copious \hzo\ emission within several to
 (in Orion-KL) several tens of arcseconds from the exciting source.
 This \hzo\ maser/accretion shock scenario is certainly speculative and
 has one obvious weakness: Given the dimensions of the continuum
 emission region only one, but not both of the regions with strong
 redshifted \hzo\ emission regions can be coincident with the
 continuum. This assumes that the continuum source has the same size at
 22.2 GHz, where the emission is optically thick, as at 43.3 GHz, where
 it is becoming optically thin. Moreover, there is the detection of
 blueshifted emission in the Effelsberg spectrum 1000 AU from the star, 
 which is clearly
 inconsistent with a pure accretion shock scenario, necessitating 
 the assumption of an additional outflow component.  Actually, 
not only the blueshifted
\hzo\ masers must be outflowing, but also some of the redshifted
   ones, else a very high central mass is required.
 %This is not necessarily the case.

 Finally, we would like to mention that Fiebig (1997) modeled the water
 emission emission observed toward the FU Orionis star L1287 as clumps
 falling on the accretion disk, trying to explain the observed velocity
 distribution. We have not explored whether our \hzo\ observations
 could be explained by a variation of his model.

 \subsection{Interstellar water masers {\it not} associated with bipolar outflows}

 It is clear from the observed wide radial velocity ranges, elongated
 emission distributions, and, most persuasively, measured proper
 motions that at least many, if not the majority of interstellar \hzo\
 masers are formed in, mostly bipolar, outflows (Reid \&\ Moran 1988).
 Particularly collimated examples include W49 and W3(OH)-\hzo (Gwinn et al.
 1992; Alcolea et al. 1993). A model for masers with this morphology was 
 presented by Mac Low \&\ Elitzur (1992) and Mac Low et al. (1994).

 In many cases where masers in both OH and \hzo\ were mapped, the
 masers in the two species are frequently found in the same regions on
 a, say, 0.1 pc scale. However, on smaller scales one mostly finds
 distinctly different distributions for the two species. Moreover, in
 most cases OH maser emission covers a significantly smaller velocity
 range as \hzo\ emission, indicating that it emerges from the slowly (a
 few \kms) outmoving or infalling envelopes of young stars; in some
 cases, e.g. W3(OH), from just outside the ionization/shock front of an
 UCHII region (Reid et al. 1980)\footnote{While this picture for
 interstellar OH masers probably holds for the majority of the stronger
 sources, recently weak OH maser emission has also been identified
 offset from the UCHII/OH maser source W3(OH), whose position places it
 at the working surface of the \hzo\ maser outflow that originates from
 a deeply embedded protostar neighboring the former region (Argon et
 al. 2003).}

 Given the completely different pumping requirements of \hzo\ and OH
 masers ($n\sim 10^9$ \ccm/$T\sim400$ K; Elitzur et al. 1989) and
 ($\sim10^7$ \ccm/$\sim150$ K; Cesaroni \&\ Walmsley 1991),
 respectively, it is clear that both masers arise in quite different
 gas volumes ($n$ and $T$ are density and kinetic temperature,
 respectively).  The apparently contradictory result that \hzo\ masers
 occur further away from their exciting sources than OH masers is
 explained by the fact that they arise from hot, compressed postshock
 material. Also, cases were OH and \hzo\ masers appear on the same spot
 in the sky may be chance projections.

 That the CRL 2136 maser is located so close (closer than the OH
 masers) to the exciting source, as well as its velocity structure,
 indicates that we are dealing with a maser that is {\it not}
 associated with an outflow, but, as discussed above, possibly with
 infall. This interpretation is corroborated by the absence of
 high-velocity \hzo\ emission.  Could  a whole class of such
 \hzo\ masers exist, which are  clearly identified on observational 
 grounds, but whose ``distinction'' from  outflow-associated masers
 has not yet been recognized?  We consider
 this to be entirely conceivable. Inspection of the OH and \hzo\ maps
 of Forster \&\ Caswell (1999) reveals a number of candidate sources
 that sometimes, although not always, are associated with weak
 contiunuum emission.

 Another source showing a very similar relationship between \hzo\
 masers and a weak, compact, and elongated radio continuum source as
 CRL 2136 is AFGL 2591 (Trinidad et al. 2003), an infrared source of
 comparable luminosity to CRL 2136 [$2\times10^4$ \Lsun\ (assuming 
 $D=1$ kpc)
 compared to
 CRL 2136's $5\times10^4$ \Lsun)]. Here the \hzo\ masers are
 concentrated in a $~\sim 60$ AU region.

 We also mention the remarkable \hzo\ maser distribution mapped  
 by Torrelles et al. (2001) toward Cepheus AHW2, which to 
 great accuracy traces part of a
 circular arc (of radius 62 AU), indicating spherical, episodic ejection.
 Recently, this arc was found to be expanding (Gallimore et al. 2003).
 Other regions with extremely compact \hzo\ maser distributions (more
 compact than CRL 2136's $600\times1000$ AU) are NGC 2071-IRS3
 (40 AU possibly in a disk; Torrelles et al. 1998), 
 and W75 N(B) (150 AU; Torrelles et al. 1997).

 Finally, we mention that in Orion-KL, Genzel et al. (1980) identify,
 in addition to a high- and a low-velocity \hzo\ outflow, the so-called
 ``shell'' masers\footnote{These masers are called ``shell-type'' by
 Genzel et al. because their velocity structure resembles that of \hzo\
 masers in evolved star circumstellar shells. These masers are
 ``resolved out'' by Genzel et al.'s and other high-resolution VLBI
 observations, although the latter authors give Hansen (1980) as a VLBI
 reference. A good VLA A-array map of the Orion \hzo\ masers seems
 highly desirable!}, which only occur in the immediate vicinity of IRc
 2 (= source I) and have apparent maser spot sizes that are an order of
 magnitude larger than any other maser in the Orion region. These
 masers may be the archetype of the new class identified here.

 \section{Conclusions and outlook}

 Using the VLA, we have detected several weak radio continuum sources in
 the CRL 2136 region. One of these, RS 4, is, within the errors,
 coincident with IRS 1, the high-mass protostar exciting the region.
 Taking our 8.4, 14.9, and 43.3 GHz data and 86 GHz data from the
 literature, the emission, which is almost certainly free-free
 radiation, has a rising spectral index, $\alpha$
 $(S\propto\nu^\alpha)$ of 1.2 up to 43.3 GHz, which flattens at higher
 frequencies.  The continuum emission might be arising from a bipolar
 jet, as modeled by Tan (2003), and it seems highly desirable to apply
 his model to the region discussed here.

 Water maser emission was found from a very confined region of size
 $0\as3\times0\as5$ ($600\times1000$ AU) with its centroid coincident
 with RS 4. All of the  strong emission is redshifted
 relative to the systemic velocity by up to 4 \kms. The strongest
 emission arises from a single feature (at $v_{\rm peak} = 27.1$ \kms),
 which appears to have been at the same velocity (to within
 $\approx\pm0.1$ \kms) for a period of at least 9 years.  Given the
 observed redshift, it is interesting to speculate that the
 water-containing gas giving rise to the strong emission is falling
 onto the central protostar and is boosted by amplification of the
 background continuum emission. Given the measured size of the
 continuum emission region ($0\as029\times0\as003$) this can only be
 true for part of the strong emission, which is arising from two
 compact regions 0\as08 apart.
 %The absence of strong blueshifted emission is naturally explained
 %by the fact that the intervening free-free emission is optically thick.
 %The spatial distribution of the  extremely weak ($S<0.1$ Jy) 
 %emission at and below the systemic
 %velocity ($S<0.1$ Jy), as well as that of the weak highest velocity
 %emission (\vlsr $> 30$ \kms, $S<0.5$ Jy)  may be emitted from material
 %at located at the edge of the continuum emission; 
 {\it Simultaneous} high ($<0\as1$) resolution VLA observations of the
\hzo\ maser emission and the 22 GHz continuum emission will provide a
detailed picture of the relationship between the two phenomena 
(reducing the cross-registration uncertainty to a few milli-arcseconds)
and
certainly prove or disprove the accretion shock scenario.  Using the
\hzo\ maser as a phase reference (Reid \&\ Menten 1990, 1997) will allow
very high quality imaging.

The accretion shock seems a  natural environment for the
production of the  \hzo\ maser emission, which requires temperatures
around 400 K and densities between $10^8$ and $10^9$ \ccm.
Modeling efforts should explain the velocity stability of
the strong maser feature.
% for which we derive an upper limit on its acceleration of
%$10^{-9}$ km~s$^{-2}$.
The model of free-free emission from 
accretion shocks by Neufeld \&\ Hollenbach (1996) under-predicts 
the observed radio continuum by several orders of magnitude.

Finally, we speculate that the CRL 2136 \hzo\ masers belong to a
not yet identified class of \hzo\ masers that are in the closest
vicinity of the protostar and do not partake in outflows, but
possibly are part of the infalling material. The prototype
of this class are the ``shell-type'' masers in Orion-KL.

\acknowledgements{We are grateful to Christian Henkel for taking the
Effelsberg spectrum and to Jonathan Tan, Mark Reid, and
Malcolm Walmsley for
comments on the manuscript. We thank Joel Kastner for providing 
information on his old maser data. An anonymous referee provided valuable
comments that led to a significantly improved paper.
The 100-m telescope at Effelsberg is
operated by the Max-Planck-Institut f\"ur Radioastronomie (MPIfR).

\clearpage
\newpage

\newpage

\end{document}